\documentclass[conference]{IEEEtran}
\IEEEoverridecommandlockouts

\usepackage[cmex10]{amsmath}
\usepackage{array}
\usepackage{amssymb}
\usepackage{bussproofs}
\usepackage{cmll}
\usepackage{fixltx2e}
\usepackage{url}
\usepackage{proof}
\usepackage{stmaryrd}
\usepackage{graphicx}
\usepackage{xspace}
\usepackage[all]{xy}
\usepackage{listings}
\usepackage{multicol}

\newtheorem{theorem}{Theorem}[section]
\newtheorem{lemma}[theorem]{Lemma}
\newtheorem{definition}[theorem]{Definition}
\newtheorem{proposition}[theorem]{Proposition}

\newcommand{\omitnow}[1]{}
\newcommand{\bfalse}{\texttt{\textbf{false}}\xspace}
\newcommand{\btrue}{\texttt{\textbf{true}}\xspace}

\newcommand{\arrow}[1]{#1}

\newcommand{\mat}{\Leftrightarrow}

\newcommand{\alt}{~|~}

\newcommand{\ket}[1]{|#1\rangle}
\newcommand{\singleton}[1]{\{#1\}}
\newcommand{\meas}[1]{\textit{meas}~#1}

\newcommand{\ip}[2]{\langle #1 \alt #2 \rangle}
\newcommand{\op}[2]{|#1\rangle\langle #2|}

\newcommand{\scalarZ}{\texttt{F}}
\newcommand{\scalarO}{\texttt{T}}
\newcommand{\tte}{\texttt{tt}}
\newcommand{\ffe}{\texttt{ff}}

\newcommand{\scalarPlus}{\ensuremath{\underline{\vee}}}
\newcommand{\scalarTimes}{\ensuremath{\wedge}}

\newcommand{\lete}[2]{\textbf{let}~#1~\textbf{in}~#2}

\newcommand{\DQT}{\ensuremath{\mathrm{DQT}_{\mathsf{2}}}}
\newcommand{\DQC}{\ensuremath{\mathrm{DQC}_{\mathsf{2}}}}
\newcommand{\DQCn}{\ensuremath{\mathrm{DQC}_{n}}}

\begin{document}
\title{Quantum Computing over Finite Fields: \\
Reversible Relational Programming with \\
Exclusive Disjunctions}

% \author{
% \IEEEauthorblockN{
%   Roshan P. James\IEEEauthorrefmark{1}, 
%   Gerardo Ortiz\IEEEauthorrefmark{2}, and
%   Amr Sabry\IEEEauthorrefmark{1}\thanks{Partially funded by IU's Office of the Vice Provost for Research through its Faculty Research Support Program.}}
% \IEEEauthorblockA{\IEEEauthorrefmark{1}School of Informatics and Computing\\
%   Indiana University\\
%   Email: \{rpjames,sabry\}@indiana.edu} 
% \and
% \IEEEauthorblockA{\IEEEauthorrefmark{2}Department of Physics\\
%   Indiana University\\
%   Email: ortizg@indiana.edu}
% }

\author{ 
  \IEEEauthorblockN{Roshan P. James}
  \IEEEauthorblockA{School of Informatics and Computing\\
    Indiana University\\
    Email: rpjames@indiana.edu} \and 
  \IEEEauthorblockN{Gerardo Ortiz}
  \IEEEauthorblockA{Department of Physics\\
    Indiana University\\
    Email: ortizg@indiana.edu} \and 
  \IEEEauthorblockN{Amr Sabry}
  \IEEEauthorblockA{School of Informatics and Computing\\
    Indiana University\\
    Email: sabry@indiana.edu} 
  \thanks{Partially funded by IU's Office of the Vice Provost for
    Research through its Faculty Research Support Program.}
}

\maketitle

%%%%%%%%%%%%%%%%%%%%%%%%%%%%%%%%%%%%%%%%%%%%%%%%%%%%%%%%%%%%%%%%%%%%%%%%%%%%%
\begin{abstract}
In recent work, Benjamin Schumacher and Michael~D. Westmoreland investigate a
version of quantum mechanics which they call \emph{modal quantum theory} but
which we prefer to call \emph{discrete quantum theory}.  This theory is
obtained by instantiating the mathematical framework of Hilbert spaces with a
finite field instead of the field of complex numbers. This instantiation
collapses much the structure of actual quantum mechanics but retains several
of its distinguishing characteristics including the notions of superposition,
interference, and entanglement. Furthermore, discrete quantum theory excludes
local hidden variable models, has a no-cloning theorem, and can express
natural counterparts of quantum information protocols such as superdense
coding and teleportation.

Our first result is to distill a model of discrete quantum computing from
this quantum theory. The model is expressed using a monadic metalanguage
built on top of a universal reversible language for finite computations, and
hence is directly implementable in a language like Haskell. In addition to
superpositions and invertible linear maps, the model includes conventional
programming constructs including pairs, sums, higher-order functions, and
recursion. Our second result is to relate this programming model to
relational programming, e.g., a pure version of Prolog over finite
relations. Surprisingly discrete quantum computing is identical to
conventional logic programming except for a small twist that is responsible
for all the ``quantum-ness.'' The twist occurs when merging sets of answers
computed by several alternatives: the answers are combined using an
\emph{exclusive} version of logical disjunction. In other words, the two
branches of a choice junction exhibit an \emph{interference} effect: an
answer is produced from the junction if it occurs in one or the other branch
but not both. 
\end{abstract}

%%%%%%%%%%%%%%%%%%%%%%%%%%%%%%%%%%%%%%%%%%%%%%%%%%%%%%%%%%%%%%%%%%%%%%%%%%%%%
\section{The Result, Informally} 
\label{sec:rel}

Consider this Prolog program:

\lstset{
        language=Prolog,
 morekeywords={true, false},
	keywordstyle=\bfseries\ttfamily,
	identifierstyle=\ttfamily,
	commentstyle=\ttfamily,
	stringstyle=\ttfamily,
	showstringspaces=false,
	basicstyle=\scriptsize\sffamily,
	tabsize=2,
	breaklines=true,
	prebreak = \raisebox{0ex}[0ex][0ex]{\ensuremath{\hookleftarrow}},
	breakatwhitespace=false,
	aboveskip={1.2\baselineskip},
        columns=fixed,
        extendedchars=true,
}
\begin{lstlisting}
r(false,false).
r(false,true).
r(true,false).

q(X) :- r(false,X).
q(X) :- r(true,X).
 \end{lstlisting}
%subcode source qmll.tex:270

The program starts with three facts about a relation \verb|r| and then
defines a rule \verb|q| such that \verb|q(X)| is true if either of the two
clauses can be satisfied. Executing the query \verb|q(X).| returns three
answers:

\lstset{
        language=Prolog,
 morekeywords={true, false},
	keywordstyle=\bfseries\ttfamily,
	identifierstyle=\ttfamily,
	commentstyle=\ttfamily,
	stringstyle=\ttfamily,
	showstringspaces=false,
	basicstyle=\scriptsize\sffamily,
	tabsize=2,
	breaklines=true,
	prebreak = \raisebox{0ex}[0ex][0ex]{\ensuremath{\hookleftarrow}},
	breakatwhitespace=false,
	aboveskip={1.2\baselineskip},
        columns=fixed,
        extendedchars=true,
}
\begin{lstlisting}
X = false ;
X = true ; 
X = false.
 \end{lstlisting}
%subcode source qmll.tex:281

Now consider the same example expressed in the Discrete Quantum Theory over
the field of booleans (\DQT) recently developed by Schumacher and
Westmoreland~\cite{modalqm}:

\[\begin{array}{rcl}
r\ket{0} &=& \ket{0} + \ket{1} \\
r\ket{1} &=& \ket{0} \\
\\
q &=& r\ket{0} + r\ket{1}
\end{array}\]

The relation \verb|r| which could relate \bfalse to either \bfalse or \btrue
is expressed as a quantum gate $r$ that maps $\ket{0}$ to a superposition of
$\ket{0}$ and $\ket{1}$. The rule \verb|q| corresponds to a vector~$q$
produced by taking the superposition of the two possible alternatives.

In contrast to the Prolog program, measuring the vector $q$ in the standard
basis $\{\ket{0},\ket{1}\}$ is guaranteed to \emph{always} return~$\ket{1}$:
the answer $\ket{0}$ could never be produced! To understand why, we calculate
the result of $q$ as follows:
\[
q = r\ket{0} + r\ket{1} = (\ket{0}+\ket{1}) + \ket{0} = \ket{1}
\]
In the last step the two intermediate answers $\ket{0}$ \emph{interfere}
destructively with each other and are canceled. The details are explained in
Section~\ref{sec:dqt}.

This simple example captures the essence of our result which can be
informally stated as follows. \emph{The model of computation inherent in
\DQT, \DQC, is a relational programming model in which any value that appears
an even number of times in disjunctions disappears.}
Section~\ref{sec:sdcoding} shows a more significant example that implements
the superdense coding quantum protocol in both \DQC\ and Prolog. Removing all
answers that appear an even number of times in the Prolog execution is indeed
consistent with the quantum protocol.

%%%%%%%%%%%%%%%%%%%%%%%%%%%%%%%%%%%%%%%%%%%%%%%%%%%%%%%%%%%%%%%%%%%%%%%%%%%%%
\section{The Result, Formally} 
\label{sec:towards}

The pioneering research programmes of Abramsky and
Coecke~et~al.~\cite{1021878,categoricalQM} and of
Selinger~\cite{DBLP:journals/mscs/Selinger04,Selinger-dagger}
have established that quantum computing (QC) can be modeled using
\emph{dagger compact closed categories}. The traditional mathematical model
of QC --- the category $\mathrm{FDHilb}$ of finite dimensional Hilbert spaces
and linear maps --- is a prime example of such categories, \emph{and so is
the category $\mathrm{Rel}_{\,\cup}$ of sets and relations}. Our result,
formally stated, is that the computational structures inherent in \DQC\ can
be modeled in a non-standard category $\mathrm{Rel}_{\,\uplus}$ of sets and
relations which is isomorphic to the category $\mathrm{FDVec}_2$ of finite
dimensional vector spaces over the field of booleans and linear maps. This
category is dagger compact closed and hence possesses all the computational
structures necessary for QC. Our category differs from the standard category
of sets and relations in one aspect: it uses the \emph{exclusive union} of
sets everywhere the standard union would be used. Diagrammatically we have:

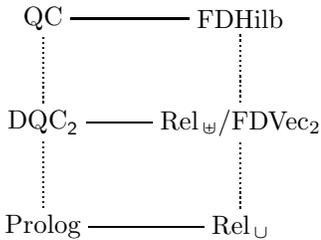
\[
\xymatrix{
  \mathrm{QC} \ar@{-}[r] \ar@{.}[d]   & \mathrm{FDHilb} \ar@{.}[d] \\
  \DQC \ar@{-}[r] \ar@{.}[d] & \mathrm{Rel}_{\,\uplus}/\mathrm{FDVec}_2 \ar@{.}[d] \\
  \mathrm{Prolog} \ar@{-}[r]             & \mathrm{Rel}_{\,\cup} 
}
\]

%%%%%%%%%%%%%%%%%%%%%%%%%%%%%%%%%%%%%%%%%%%%%%%%%%%%%%%%%%%%%%%%%%%%%%%%%%%%%
\section{Significance and Background} 

\paragraph*{Hilbert Spaces and Models of Quantum Computing} 
The traditional mathematical formulation of Quantum Mechanics (QM) is founded
on Hilbert spaces. Although there are several other more abstract
mathematical formulations of QM, the Hilbert space formalism remains the most
accepted and the most widely used such formalism~\cite{544199}. A Hilbert
space is defined to be a real or complex inner product space that is a
complete metric space with respect to the distance function induced by the
inner product. A real or complex inner product space is a vector space over
the field of real or complex numbers on which there is an inner product. In
traditional QM, the underlying field is the field of complex numbers which
serves as the space from which \emph{probability amplitudes} for quantum
events are drawn.

For the purposes of QC, the mathematical formalism of QM is typically
restricted to finite dimensional Hilbert spaces. This restriction removes one
of the infinities in the formalism but retains another --- the underlying
infinite field of complex numbers. The resulting model is clearly still a
mathematical idealization as it allows infinitesimally fine distinctions
among quantum states that might differ by vanishingly small probability
amplitudes. In contrast, any particular realization of a quantum algorithm
can only assume finite and discrete levels of representation of such
probability amplitudes. Although it is customary for (classical or quantum)
models of computations to include infinite structures of various kinds, it is
important to understand how the idealized models emerge as the limits of
finite approximations. A fundamental question that therefore motivates our
research is whether it is possible to replace the infinite field of complex
numbers by successively larger and larger finite fields to reach full QC in
the limit. More importantly, what is the computational power of these
intermediate models, \DQCn, as $n$ grows larger?

To tackle these questions, we do not work directly with the Hilbert
space formalism, however. Indeed, the mathematical formulation of QC
based on Hilbert spaces obscures many traditional computational
structures that are inherent in the physical theory. Early on, Mu and
Bird~\cite{mu-bird} showed that the simple model of QC based on
vectors and linear operators is a \emph{monad} which is the standard
way to model computational effects~\cite{77353}. In a subsequent
development, the third author and collaborators~\cite{1166040,
  DBLP:conf/sbmf/VizzottoLS09, DBLP:conf/wollic/VizzottoBS09,
  DBLP:journals/entcs/VizzottoCS08,
  DBLP:journals/entcs/AltenkirchGVS07} have established that the more
general quantum model based on density matrices and superoperators is
an instance of the mathematical concept of
\emph{arrows}~\cite{hughes:arrows} which is a generalization of
monads.  Interestingly dagger closed compact categories identified by
Abramsky et. al. also serve as models of linear logic~\cite{ll} and
various computational effects \cite{power-generic,
  Hasegawa:1997:RCS:645893.671607, Power:1997:PCN:967340.967345}.
%% TODO cite thielecke 
Hence, based on these connections, our technical contributions
are expressed using traditional computational structures and constructions:
monads, arrows, and category theory.

%%%%%%%%%%%%%%%%%%%%%%%%%%%%%%%%%%%%%%%%%%%%%%%%%%%%%%%%%%%%%%%%%%%%%%%%
\section{Discrete Quantum Theory}
\label{sec:dqt} 

In their recent work, Schumacher and Westmoreland~\cite{modalqm} argue that
much of the structure of traditional QM is maintained in the presence of
finite fields. In particular, they establish that the quantum theory based on
the finite field of booleans retains the following characteristics of QM:
\begin{itemize}
\item the notions of superposition, interference, entanglement, and mixed
states of quantum systems;
\item the time evolution of quantum systems using invertible linear
operators;
\item the complementarity of incompatible observables;
\item the exclusion of local hidden variable theories and the impossibility
of cloning quantum states; and
\item the presence of natural counterparts of quantum information protocols
such as superdense coding and teleportation.
\end{itemize}

\paragraph*{Fields} 
A field is an algebraic structure with notions of addition and multiplication
that satisfy the usual axioms.  The rationals, reals, complex numbers, and
quaternions form fields that are infinite. There are also finite fields that
satisfy the same set of axioms. Finite fields are necessarily ``cyclic.''

We fix a field $\mathbb{B}$ consisting of two scalars $\{ \scalarZ,
\scalarO\}$. The elements $\scalarZ$ and $\scalarO$ are associated
with the probabilities of quantum events, with $\scalarZ$ interpreted
as \emph{definitely no} and $\scalarO$ interpreted as \emph{possibly
  yes}.\footnote{Everything works if we switch the interpretation with
  $\scalarZ$ interpreted as \emph{possibly no} and $\scalarO$ as
  \emph{definitely yes}~\cite{DBLP:journals/corr/abs-0910-2393}. } The
field $\mathbb{B}$ comes with an addition operation~\scalarPlus\
(which in this case must be exclusive-or) and a multiplication
operation~\scalarTimes\ (which in this case must be conjunction). In
particular we have:

\[\begin{array}{rclcl@{\qquad\qquad\quad}rclcl}
\scalarZ &\scalarPlus& \scalarZ &=& \scalarZ & 
          \scalarZ &\scalarTimes& \scalarZ &=& \scalarZ \\
\scalarZ &\scalarPlus& \scalarO &=& \scalarO &
          \scalarZ &\scalarTimes& \scalarO &=& \scalarZ \\
\scalarO &\scalarPlus& \scalarZ &=& \scalarO &
          \scalarO &\scalarTimes& \scalarZ &=& \scalarZ \\
\scalarO &\scalarPlus& \scalarO &=& \scalarZ &
          \scalarO &\scalarTimes& \scalarO &=& \scalarO 
\end{array}\]

The definitions are intuitively consistent with the interpretation of
scalars as probabilities for quantum events except that it appears
strange to have $\scalarO \scalarPlus \scalarO$ be defined as
$\scalarZ$, i.e., to have a twice-possible event become
impossible. This results from the cyclic property intrinsic to finite
fields requiring the existence of an inverse to $\scalarPlus$. This
inverse must be $\scalarO$ itself which means that $\scalarO$
essentially plays both the roles of ``possible with phase~1'' and
``possible with phase -1'' and that the two occurrences cancel each
other in a superposition.

\paragraph*{Vector Spaces} 
In the vector space over the field $\mathbb{B}$, if vectors are represented
as functions $\kappa$ that map basis elements~$v$ to scalars field values,
then vector addition and scalar multiplication are defined by lifting the
field operations to vectors as follows:
\[\begin{array}{rcl}
\kappa_1 + \kappa_2 &=& \kappa' \qquad\mbox{s.t.}~
  \kappa'(v) = \kappa_1(v) ~\scalarPlus~ \kappa_2(v) \\
b * \kappa &=& \kappa' \qquad\mbox{s.t.}~
  \kappa'(v) = b ~\scalarTimes~ \kappa(v)
\end{array}\]
Scalar multiplication is uninteresting as it either leaves the vector
unchanged or returns the zero vector (denoted $\bullet$) which maps
everything to the scalar zero. Vector addition however is the key to
introducing superpositions and interference effects. 

Consider the simple case of a 1-qubit system with bases $\ket{0}$
and~$\ket{1}$. The Hilbert space framework allows us to construct an
infinite number of states for the qubit all of the form $\alpha\ket{0}
+ \beta\ket{1}$ with $\alpha$ and $\beta$ elements of the underlying
field of complex numbers and with the side condition that the state is
not the zero vector and that it has length 1, i.e., that
$|\alpha|^2+|\beta|^2=1$. Moving to a finite field immediately limits
the set of possible states as the coefficients $\alpha$ and~$\beta$
are now drawn from a finite set.  In other words, in the field
$\mathbb{B}$, there are exactly three valid states for the qubit:
$\scalarZ\ket{0} ~+~ \scalarO\ket{1}$ (which is equivalent
to~$\ket{1}$), $\scalarO\ket{0} ~+~ \scalarZ\ket{1}$ (which is
equivalent to $\ket{0}$), and $\scalarO\ket{0} ~+~ \scalarO\ket{1}$
(which we write as $\ket{+}$). The fourth possibility is the zero
vector which is not an allowed quantum state (discussed below). In a
larger field with three scalars, there would be eight possible states
for the qubit which intuitively suggests that one must ``pay'' for the
amount of desired superpositions: the larger the finite field, the
more states are present with the full Bloch sphere seemingly appearing
at the ``limit'' $n \rightarrow \infty$.

Interestingly, we can easily check that the three possible vectors for a
1-qubit state are linearly dependent with any pair of vectors expressing the
third as a superposition:
\[\begin{array}{rclcl}
\ket{0} &+& \ket{1} &=& \ket{+} \\
\ket{0} &+& \ket{+} &=& \ket{1} \\
\ket{1} &+& \ket{+} &=& \ket{0} 
\end{array}\]
\label{superp}
\noindent
In other words, other than the standard basis consisting of $\ket{0}$
and $\ket{1}$, there are just two other possible bases for this vector
space, $\singleton{\ket{1}, \ket{+}}$ and $\singleton{\ket{+},
  \ket{0}}$.

The example also shows that the cyclic structure of the field extends to the
vector space.

\paragraph*{Inner Products}
A Hilbert space comes equipped with an inner product $\ip{v_1}{v_2}$ which is
an operation that associates each pair of vectors with a complex number
scalar value that quantifies the ``closeness'' of the two vectors. The inner
product induces a norm $\sqrt{\ip{v}{v}}$ that can be thought of as the
length of vector $v$. In a finite field, we can still define an operation
$\ip{v_1}{v_2}$ which, following our interpretation of the scalars, would
need to return $\scalarZ$ if the vectors are definitely not the same and
$\scalarO$ if the vectors are possibly the same. This operation however does
\emph{not} yield an inner product, as the definition of inner products
requires that the field has characteristic 0, i.e., that the repeated
addition of $\scalarO$ to itself never reaches $\scalarZ$. As the above
paragraph shows this is not the case for $\mathbb{B}$ (nor for any finite
field for that matter) as the sum of positive elements must eventually ``wrap
around.'' In other words, if we choose to instantiate the mathematical
framework of Hilbert spaces with a finite field, we must therefore drop the
requirement for inner products and content ourselves with a plain vector
space. Furthermore, in the absence of a
notion of length, one must exclude zero
vectors.

\paragraph*{Invertible Linear Maps} 
In actual QC, the dynamic evolution of quantum states is described by unitary
transformations which preserve inner products. As discrete quantum theory
lacks inner products, the dynamic evolution of quantum states is described by
any invertible linear transformation, i.e., by any linear transformation that
is guaranteed never to produce the zero vector.

%% BAD-------------------------------
%%  f0 |0> = *,   |1> = *,   |+> = *
%%  f1 |0> = *,   |1> = |1>, |+> = |1> 
%%  f2 |0> = *,   |1> = |0>, |+> = |0> 
%%  f3 |0> = *,   |1> = |+>, |+> = |+>  
%%  f4 |0> = |1>, |1> = *,   |+> = |1>  
%%  f5 |0> = |1>, |1> = |1>, |+> = *
%%  f8 |0> = |0>, |1> = *  , |+> = |0>
%% f10 |0> = |0>, |1> = |0>, |+> = *
%% f12 |0> = |+>, |1> = *,   |+> = |+>
%% f15 |0> = |+>, |1> = |+>, |+> = *
%% GOOD------------------------------
%%  f6 |0> = |1>, |1> = |0>, |+> = |+>
%%  f7 |0> = |1>, |1> = |+>, |+> = |0>
%%  f9 |0> = |0>, |1> = |1>, |+> = |+>
%% f11 |0> = |0>, |1> = |+>, |+> = |1>
%% f13 |0> = |+>, |1> = |1>, |+> = |0>
%% f14 |0> = |+>, |1> = |0>, |+> = |1>
%% ----------------------------------

As an example, there are 16 linear (not necessarily invertible) functions in
the space of 1-qubit functions. Out of these, six are permutations on the
three 1-qubit vectors; the remaining all map one of the vectors to $\bullet$
which makes them non-invertible. Hence the space is quite impoverished
compared to the full set of 1-qubit linear transformations in the Hilbert
space. In particular, even some of the elementary unitary transformations
such as the Hadamard transformation are not expressible in that space. Indeed
the Hadamard matrix for the field of booleans maps the vector $\ket{+}$ to
the zero vector and hence is not an acceptable 1-qubit transformation.

\paragraph*{Entanglement and Superdense Coding} 
Despite the restriction to finite fields and the drastic reduction in the
state space of qubits and their transformations, the theory built on the
field of booleans has a definite quantum character. We present the superdense
coding example from the paper of Schumacher and Westmoreland~\cite{modalqm}
in Section~\ref{sec:sdcoding} and refer the reader to their paper for more
details.

%%%%%%%%%%%%%%%%%%%%%%%%%%%%%%%%%%%%%%%%%%%%%%%%%%%%%%%%%%%%%%%%%%%%%%%%%%%%%
\section{Outline of Technical Development}

Our aims are to (i) develop a typed programming language that pins down the
unique features of discrete quantum computation and, (ii) to relate this
language to that of conventional relational programming. Building on the work
of Abramsky and Coecke~\cite{1021878,categoricalQM} and of
Selinger~\cite{DBLP:journals/mscs/Selinger04,Selinger-dagger},
both tasks can be achieved by distilling a logic and type system from a
special category of sets and relations and by designing an appropriate
syntax. That category (and hence the language) can be related to the
conventional category of sets and relations to establish a connection with
conventional relational programming and it can be connected to discrete
quantum computation by showing that it is dagger compact closed. We proceed
in the following two stages.

\paragraph*{Classical Computation} 
Many programming models of QC start with the $\lambda$-calculus as the
underlying classical language and add quantum features on top of
it~\cite{vanTonder:2004,duncan,1079699,DBLP:journals/mscs/SelingerV06}.
% TODO cite more papers of Selinger, Arrighi, Altenkirch and others
This strategy is natural given that the $\lambda$-calculus is the canonical
classical computational model. However this strategy complicates the
development of quantum languages as it forces the languages to deal in fairly
complicated ways with the implicit duplication and erasure of information in
the classical sublanguage. A simpler strategy that loses no generality is to
build the quantum features on top of a reversible classical language: this
keeps the language simple while still enabling the $\lambda$-calculus
constructs to encoded using two explicit operators for erasure and
duplication if desired~\cite{infeffects}. Following this strategy we review
the reversible language \ensuremath{\Pi } that was recently developed by the first
and third author and use it as a foundation for the quantum language in the
remainder of the paper. The language provides the symmetrical monoidal part
of the required categorical structure.

\paragraph*{Monads and Arrows for Quantum Computation} 
We define a language \ensuremath{\mathcal{R}\Pi } by adding a layer on top of the classical
reversible language \ensuremath{\Pi } that provides sets and relations. The set
construction can be expressed as a \emph{strong monad}~\cite{complam} over
the language of classical observable values. However in order to express the
required compact closure structure in the language, the Kleisli maps of this
monad are abstracted in a first-order data type using an
arrow~\cite{hughes:arrows}. The language \ensuremath{\mathcal{R}\Pi } is expressive enough to
implement the superdense coding protocol and other quantum protocols. It
differs from a conventional relational programming language in the semantics
of the ``union'' operation. 

%%%%%%%%%%%%%%%%%%%%%%%%%%%%%%%%%%%%%%%%%%%%%%%%%%%%%%%%%%%%%%%%%%%%%%%%%%%%%
\section{Classical Computation}
\label{sec:cval}

We introduce the language \ensuremath{\Pi } for classical observable values. The
language is defined over finite types: the only ``computations'' in it are
the isomorphisms between these types. In the context of Hilbert spaces, this
language formalizes the notation used for the bases. In the categorical
context, this language provides the underlying symmetric monoidal structure.

%%%%%%%%%%%%%%%%%%%
\subsection{Bases for Vector Spaces}

In the traditional presentation of quantum computing, a Hilbert space of
dimension $d$ is spanned by a set of~$d$ mutually orthogonal vectors. For
example:
\begin{itemize}
\item a quantum system $Q_2$ composed of just 1 qubit is of dimension 2 and
is spanned by $\{ \ket{0}, \ket{1} \}$; 
\item a quantum system $Q_4$ composed of 2 qubits is of dimension 4 and is
spanned by $\{ \ket{00}, \ket{01}, \ket{10}, \ket{11} \}$.
\end{itemize}
The Hilbert space for the composition of two systems is spanned by basis
vectors of the form $\ket{m ~\cdot~ n}$ where~$m$ is a basis vector of the
first system and $n$ is a basis vector of the second system and the
operation~$(~\cdot~)$ is the composition of the two labels. For example:
\begin{itemize}
\item the composition of $Q_2$ and $Q_4$ (written $Q_2 \otimes Q_4$) is of
dimension 8 and is spanned by:
\[ 
\{ \ket{0 \cdot 00}, \ket{0 \cdot 01}, \ket{0 \cdot 10}, \ket{0 \cdot 11}, 
   \ket{1 \cdot 00}, \ket{1 \cdot 01}, \ket{1 \cdot 10}, \ket{1 \cdot 11} \}
\]
\item the composition of $Q_4$ and $Q_2$ (written $Q_4 \otimes Q_2$) is also
of dimension 8 and is spanned by:
\[ 
\{ \ket{00 \cdot 0}, \ket{00 \cdot 1}, \ket{01 \cdot 0}, \ket{01 \cdot 1}, 
   \ket{10 \cdot 0}, \ket{10 \cdot 1}, \ket{11 \cdot 0}, \ket{11 \cdot 1} \}
\]
\end{itemize}
Clearly the compositions $Q_2 \otimes Q_4$ and $Q_4 \otimes Q_2$ are
isomorphic and the distinction between them is typically blurred.

%%%%%%%%%%%%%
\subsection{Syntax and Type System} 

Instead of using bits as the labels for the bases, we use structured finite
types built using the empty type, the unit type, the sum type, and the
product type. Furthermore instead of silently identifying systems like $Q_2
\otimes Q_4$ and $Q_4 \otimes Q_2$, we include an explicit set of operators
between these finite types to witness the isomorphisms. Specifically, we have
the following language of classical observable types~$b$ and values~$v$:
% %subcode{bnf} include main
% % classical observable types, b ::= 0 | 1 | b+b 
% %   &|& b*b | bool 
% % classical observable values, v ::= () | left v | right v 
% %   &|& (v, v) | true | false

\[\begin{array}{rclr}
 b &::= & 0 ~|~ 1 ~|~ b+b ~|~ b\times b ~|~ \mathit{bool}  \\
 v &::= & () ~|~ \mathit{left} ~v ~|~ \mathit{right} ~v ~|~ (v, v) ~|~ \scalarO  ~|~ \scalarZ  \\
 \end{array}\]
%subcode source qmll.tex:695

\noindent The types are finite as they will be used to define the dimensions
of our vector spaces. The type \ensuremath{0} is the empty type containing no
inhabitants. The type \ensuremath{1} has exactly one inhabitant called
\ensuremath{()}. The type $b_1+b_2$ is the disjoint union of~$b_1$ and~$b_2$
whose elements are appropriately tagged values from either type. The type
$b_1*b_2$ is the type of ordered pairs whose elements are coming from~$b_1$
and~$b_2$ respectively. Although the type of booleans is expressible as
\ensuremath{1+1} we add it as a primitive type because it corresponds to the type of
scalar field values which play an important role in the language. The type
system is summarized below:
$$
\infer{ \vdash () : 1}{
	 ~
}
\quad
\infer{ \vdash \mathit{left} ~v : b_1 + b_2}{
	 \vdash v : b_1
}
\quad
\infer{ \vdash \mathit{right} ~v : b_1 + b_2}{
	 \vdash v : b_2
}
$$
$$
\infer{ \vdash (v_1, v_2) : b_1 \times b_2}{
	 \vdash v_1 : b_1
	&
	 \vdash v_2 : b_2
}
\quad
\infer{ \vdash \scalarO  : \mathit{bool} }{
	 ~
}
\quad
\infer{ \vdash \scalarZ  : \mathit{bool} }{
	 ~
}
$$
%subcode source qmll.tex:725

The following isomorphisms are sound and complete for the given finite
types~\cite{964008}:
\[\begin{array}{rclr}
 b &\leftrightarrow & b \\
 \\
 0 + b&\leftrightarrow & b \\
 b_1 + b_2 &\leftrightarrow & b_2 + b_1 \\
 b_1 + (b_2 + b_3) &\leftrightarrow & (b_1 + b_2) + b_3 \\
 \\
 1 \times b &\leftrightarrow & b \\
 b_1 \times b_2 &\leftrightarrow & b_2 \times b_1 \\
 b_1 \times (b_2 \times b_3) &\leftrightarrow & (b_1 \times b_2) \times b_3 \\
 \\
 0 \times b &\leftrightarrow & 0 \\
 (b_1 + b_2) \times b_3 &\leftrightarrow & (b_1 \times b_3) + (b_2 \times b_3) \\
 \\
 \mathit{bool}  &\leftrightarrow & 1+1 \\
 \end{array}\]
%subcode source qmll.tex:743

$$
\infer{ b_1 \leftrightarrow b_3}{
	 b_1 \leftrightarrow b_2
	&
	 b_2 \leftrightarrow b_3
}
$$
$$
\infer{ (b_1 + b_2) \leftrightarrow (b_3 + b_4)}{
	 b_1 \leftrightarrow b_3
	&
	 b_2 \leftrightarrow b_4
}
\quad
\infer{ (b_1 \times b_2) \leftrightarrow (b_3 \times b_4)}{
	 b_1 \leftrightarrow b_3
	&
	 b_2 \leftrightarrow b_4
}
$$
%subcode source qmll.tex:756

We introduce primitive operators corresponding to the left-to-right and
right-to-left reading of each isomorphism. We gather these operators into the
table below.
\[\begin{array}{rclr}
 \mathit{id} :& b \leftrightarrow b &: \mathit{id} \\
 \curlyeqsucc_{+} :& 0 + b \leftrightarrow b &: \curlyeqprec_{+} \\
 \times_{+} :& b_1 + b_2 \leftrightarrow b_2 + b_1 &: \times_{+} \\
 \gtrless_{+} :& b_1 + (b_2 + b_3) \leftrightarrow (b_1 + b_2) + b_3 &: \lessgtr_{+} \\
 \curlyeqsucc_{\times } :& 1 \times b \leftrightarrow b &: \curlyeqprec_{\times } \\
 \times_{\times } :& b_1 \times b_2 \leftrightarrow b_2 \times b_1 &: \times_{\times } \\
 \gtrless_{\times } :& b_1 \times (b_2 \times b_3) \leftrightarrow (b_1 \times b_2) \times b_3 &: \lessgtr_{\times } \\
 \Yleft_{0} :& 0 \times b \leftrightarrow 0 &: \Yright_{0} \\
 \Yleft  :& (b_1 + b_2) \times b_3 \leftrightarrow (b_1 \times b_3) + (b_2 \times b_3) &: \Yright  \\
 \inplus  :& \mathit{bool}  \leftrightarrow 1+1 &: \niplus  \\
 \end{array}\]
%subcode source qmll.tex:771
Each line of this table is to be read as the definition of two operators. For
example corresponding to the identity of \ensuremath{\times} isomorphism we have the two
operators \ensuremath{\curlyeqsucc_{\times } : 1 \times b \leftrightarrow b} and \ensuremath{\curlyeqprec_{\times } : b \leftrightarrow 1 \times b}.

Now that we have primitive operators we need some means of composing
them. We construct the composition combinators out of the closure
conditions for isomorphisms. Thus we have one sequential composition
combinator~\ensuremath{\circ} and two parallel composition combinators, one for
sums~\ensuremath{\oplus} and one for pairs~\ensuremath{\otimes}.
$$
\infer{ (c_1\circ c_2) : b_1 \leftrightarrow b_3}{
	 c_1 : b_1 \leftrightarrow b_2
	&
	 c_2 : b_2 \leftrightarrow b_3
}
\quad
\infer{ (c_1 \oplus c_2) : (b_1 + b_2) \leftrightarrow (b_3 + b_4)}{
	 c_1 : b_1 \leftrightarrow b_3
	&
	 c_2 : b_2 \leftrightarrow b_4
}
$$
$$
\infer{ (c_1 \otimes c_2) : (b_1 \times b_2) \leftrightarrow (b_3 \times b_4)}{
	 c_1 : b_1 \leftrightarrow b_3
	&
	 c_2 : b_2 \leftrightarrow b_4
}
$$
%subcode source qmll.tex:792

\begin{definition}{(Syntax of \ensuremath{\Pi })}
\label{def:langRev}
We collect our combinators, types and values to get the
definition of our language for isomorphisms, which we will refer to as
\ensuremath{\Pi }:
\[\begin{array}{rclr}
 b &::= & 0 ~|~ 1 ~|~ b+b ~|~ b\times b ~|~ \mathit{bool}  \\
 v &::= & () ~|~ \mathit{left} ~v ~|~ \mathit{right} ~v ~|~ (v,v) ~|~ \scalarZ  ~|~ \scalarO  \\
 \\
 t &::= & b \leftrightarrow b \\
 \mathit{iso} &::= & \times_{+} ~|~ \gtrless_{+} ~|~ \lessgtr_{+} ~|~ \curlyeqsucc_{+} ~|~ \curlyeqprec_{+} \\
 & ~|~ & \times_{\times } ~|~ \gtrless_{\times } ~|~ \lessgtr_{\times } ~|~ \curlyeqsucc_{\times } ~|~ \curlyeqprec_{\times } \\
 & ~|~ & \Yleft_{0} ~|~ \Yright_{0} ~|~ \Yleft  ~|~ \Yright  ~|~ \mathit{id} ~|~ \inplus  ~|~ \niplus  \\
 c &::= & \mathit{iso} ~|~ c \circ c ~|~ c \otimes c ~|~ c \oplus c \\
 \end{array}\]
%subcode source qmll.tex:807
\end{definition}

Given a program \ensuremath{c : b_1 \leftrightarrow b_2} in \ensuremath{\Pi }, we can run it either
in the forward direction by applying it to a value \ensuremath{v_1 : b_1} or run
it in the opposite direction by applying it to a value \ensuremath{v_2 : b_2}. 
The forward \ensuremath{c ~v_1 \mapsto v_2} transitions are given below:
\[\begin{array}{clcl }
 \mathit{id} & v &\mapsto & v \\
 \curlyeqsucc_{+} & (\mathit{right} ~v) &\mapsto & v \\
 \curlyeqprec_{+} & v &\mapsto & \mathit{right} ~v \\
 \times_{+} & (\mathit{left} ~v) &\mapsto & \mathit{right} ~v \\
 \times_{+} & (\mathit{right} ~v) &\mapsto & \mathit{left} ~v \\
 \gtrless_{+} & (\mathit{left} ~v_1) &\mapsto & \mathit{left} ~(\mathit{left} ~v_1) \\
 \gtrless_{+} & (\mathit{right} ~(\mathit{left} ~v_2)) &\mapsto & \mathit{left} ~(\mathit{right} ~v_2) \\
 \gtrless_{+} & (\mathit{right} ~(\mathit{right} ~v_3)) &\mapsto & \mathit{right} ~v_3 \\
 \lessgtr_{+} & (\mathit{left} ~(\mathit{left} ~v_1)) &\mapsto & \mathit{left} ~v_1 \\
 \lessgtr_{+} & (\mathit{left} ~(\mathit{right} ~v_2)) &\mapsto & \mathit{right} ~(\mathit{left} ~v_2) \\
 \lessgtr_{+} & (\mathit{right} ~v_3) &\mapsto & \mathit{right} ~(\mathit{right} ~v_3) \\
 \curlyeqsucc_{\times } & ((), v) &\mapsto & v \\
 \curlyeqprec_{\times } & v &\mapsto & ((), v) \\
 \times_{\times } & (v_1, v_2) &\mapsto & (v_2, v_1) \\
 \gtrless_{\times } & (v_1, (v_2, v_3)) &\mapsto & ((v_1, v_2), v_3) \\
 \lessgtr_{\times } & ((v_1, v_2), v_3) &\mapsto & (v_1, (v_2, v_3)) \\
 \Yleft  & (\mathit{left} ~v_1, v_3) &\mapsto & \mathit{left} ~(v_1, v_3) \\
 \Yleft  & (\mathit{right} ~v_2, v_3) &\mapsto & \mathit{right} ~(v_2, v_3) \\
 \Yright  & (\mathit{left} ~(v_1, v_3)) &\mapsto & (\mathit{left} ~v_1, v_3) \\
 \Yright  & (\mathit{right} ~(v_2, v_3)) &\mapsto & (\mathit{right} ~v_2, v_3) \\
 \inplus  & \scalarO  &\mapsto & \mathit{left} ~() \\
 \inplus  & \scalarZ  &\mapsto & \mathit{right} ~() \\
 \niplus  & (\mathit{left} ~()) &\mapsto & \scalarO  \\
 \niplus  & (\mathit{right} ~()) &\mapsto & \scalarZ  \\
 \end{array}\]
%subcode source qmll.tex:839

Since there are no values that have the type \ensuremath{0}, the reductions for
the combinators \ensuremath{\curlyeqsucc_{+}}, \ensuremath{\curlyeqprec_{+}}, \ensuremath{\Yleft_{0}} and \ensuremath{\Yright_{0}}
omit the impossible cases. The
semantics of the other combinators is straightforward. 
Composition combinators are defined as follows:
$$
\infer{ (c_1 \oplus c_2) ~(\mathit{left} ~v_1) \mapsto \mathit{left} ~v_2}{
	 c_1 ~v_1 \mapsto v_2
}
$$
$$
\infer{ (c_1 \oplus c_2) ~(\mathit{right} ~v_1) \mapsto \mathit{right} ~v_2}{
	 c_2 ~v_1 \mapsto v_2
}
$$
$$
\infer{ (c_1 \otimes c_2) ~(v_1, v_2) \mapsto (v_3, v_4)}{
	 c_1 ~v_1 \mapsto v_3
	&
	 c_2 ~v_2 \mapsto v_4
}
\quad
\infer{ (c_1\circ c_2) ~v_1 \mapsto v_2}{
	 c_1 ~v_1 \mapsto v
	&
	 c_2 ~v \mapsto v_2
}
$$
%subcode source qmll.tex:859

For the inverse direction, it is a simple matter to establish the
following property. 

\begin{proposition}
\label{existence-of-inverses}
For every \ensuremath{\Pi } program \ensuremath{c}, such that \ensuremath{c ~v \mapsto v'}, we can
construct its adjoint \ensuremath{c{\mathit{^\dagger }}} such that \ensuremath{c{\mathit{^\dagger }} v' \mapsto v}.
\end{proposition}
\begin{IEEEproof}
  We can construct the required \ensuremath{c^{\dagger}} by replacing every
  primitive isomorphism with its dual. For sequential composition we have
  \ensuremath{ (c_1 \circ c_2)^{\dagger} = c_2^{\dagger} \circ c_1^{\dagger}}
  and for parallel composition we have \ensuremath{ (c_1 \otimes
  c_2)^{\dagger} = c_1^{\dagger} \otimes c_2^{\dagger}} and \ensuremath{(c_1
  \oplus c_2)^{\dagger} = c_1^{\dagger} \oplus c_2^{\dagger}}.
\end{IEEEproof}

%%%%%%%%%%%%%%%%%%%%%%%%%%%%%
\subsection{Dagger Symmetrical Monoidal Structure} 
\label{examples}

The language \ensuremath{\Pi } is rich enough to express the required dagger
symmetric monoidal (but not the compact close structure) structure. The proof
for this is straightforward and we content ourselves with a brief
outline. The language \ensuremath{\Pi } can be interpreted as category whose
objects are the base types~\ensuremath{b} and whose morphisms the
combinators~\ensuremath{c}. Identity morphisms are given by \ensuremath{\mathit{id}} and associativity
follows from the composition of isomorphisms, thus establishing the required
properties for a category.

Further \ensuremath{\Pi } is a monoidal category \ensuremath{(b, \times , 1)} where \ensuremath{\times}
is the tensor operation, \ensuremath{1} is the identity object and the required
natural transforms $\alpha$, $\lambda$ and $\rho$ are given by
\ensuremath{\lessgtr_{\times }}, \ensuremath{\curlyeqsucc_{\times }} and \ensuremath{\times_{\times } \circ \curlyeqsucc_{\times }} respectively. The
required ``pentagon'' and ``triangle'' axioms follow from the
definitions of the these isomorphisms.

Braiding is provided by \ensuremath{\times_{\times }} at the appropriate types and it satisfies
the ``hexagon'' axioms. Further that \ensuremath{\times_{\times }} at \ensuremath{b_1 \times b_2} is the same as
\ensuremath{\times_{\times } ^{-1}} at \ensuremath{b_2\times b_1} establishing symmetry.

Finally, proposition \ref{existence-of-inverses} tells us that for every
morphism~\ensuremath{c} there exists its adjoint $c^{\dagger}$ establishing that
\ensuremath{\Pi } is a dagger symmetric monoidal category.

%%%%%%%%%%%%%%%%%%%%%%%%%%%%%%%%%%%%%%%%%%%%%%%%%%%%%%%%%%%%%%%%%%%%%%%%%%%%%
\section{The Language \ensuremath{\mathcal{R}\Pi } } 
\label{sec:core}
\label{sec:extensions}

The core language for discrete quantum computing is obtained by adding a
layer on top of the reversible core presented in the previous section. The
additional layer is that of vectors and linear maps or equivalently that of
sets and relations.

\begin{definition}{(Syntax of Core \ensuremath{\mathcal{R}\Pi }) }
\label{def:coreLang}
\[\begin{array}{rclr}
 b &::= & ... ~|~ S ~b ~|~ b ~R ~b \\
 v &::= & ... ~|~ s \\
 s &::= & \emptyset  ~|~ \singleton {v} ~|~ s ~\uplus  ~s \\
 r &::= & \arrow{arr}  ~\mathit{iso} ~|~ r \,\arrow{\ggg\xspace}\, r ~|~ \mathit{second} ~r \\
 & ~|~ & \mathit{strength} ~|~ \mathit{state} ~s ~|~ \eta _b ~|~ \varepsilon  \\
 \end{array}\]
%subcode source qmll.tex:923
\end{definition}

The language \ensuremath{\mathcal{R}\Pi } extends \ensuremath{\Pi } as follows. The set of types is
extended with the type \ensuremath{S ~b} of sets of values and the type \ensuremath{b_1 ~R ~b_2} of
relations between sets of values of type~\ensuremath{b_1} and sets of values of
type~\ensuremath{b_2}. The set of values is extended with sets~\ensuremath{s} which can either
be the empty set \ensuremath{\emptyset }, a singleton set \ensuremath{\singleton {v}}, or the
\emph{exclusive union} \ensuremath{s_1 ~\uplus  ~s_2}. The exclusive union of sets
\ensuremath{s_1} and \ensuremath{s_2} is the union of all the elements that are not in
common. This union reflects the cyclic nature of the underlying finite field
as was explained in Section~\ref{sec:dqt}. In more detail, we saw for example
that adding the vector $\scalarZ\ket{0} ~+~ \scalarO\ket{1}$ to the vector
$\scalarO\ket{0} ~+~ \scalarO\ket{1}$ produces the vector $\scalarO\ket{0}
~+~ \scalarZ\ket{1}$ where the two occurrences of the component $\ket{1}$
canceled each other. Expressed as sets, we would have that $\{ 1 \} \uplus \{
0,1 \}$ is equal to $\{ 0 \}$.

The layer of sets corresponds to a \emph{strong monad} with the singleton set
as the unit and the folding of the exclusive union as the bind operation. The
Kleisli arrows of this monad are functions that map values to sets. In order
to express these functions themselves as relations, we provide explicit arrow
operators to construct them. The language therefore includes a separate
syntactic category of relations which is used to build the required Kleisli
maps using arrow primitives. The operator \ensuremath{\arrow{arr} } lifts any isomorphism from
the underlying reversible language to a relation on sets. The operator
\ensuremath{\,\arrow{\ggg\xspace}\,} sequences two relations and the operator \ensuremath{\mathit{second}} applies a
relation to the second component of a pair leaving the first component
unchanged. These three operators are the minimal core operators for any arrow
language. In addition, we must include enough structure to provide a compact
closed category: the operator \ensuremath{\mathit{strength}} is used to express the tensor
product of sets; the operator \ensuremath{\mathit{state}} identifies a state of type \ensuremath{S ~b}
with an arrow (relation) of type \ensuremath{1 ~R ~b}~\cite{categoricalQM}; finally the
pair of operators~$\ensuremath{\eta }_b$ and \ensuremath{\varepsilon } are the unit elimination and
introduction that are required for any compact closed category. The
operation~\ensuremath{\eta } is indexed by a type \ensuremath{b}. 

We present the type system below. The remainder of this section explains the
semantics of core \ensuremath{\mathcal{R}\Pi } formally and illustrates its use in several
examples:
$$
\infer{ \vdash \emptyset  : S ~b}{
	 ~
}
\quad
\infer{ \vdash \singleton {v} : S ~b}{
	 \vdash v : b
}
\quad
\infer{ \vdash s_1 ~\uplus  ~s_2 : S ~b}{
	 \vdash s_1 : S ~b
	&
	 \vdash s_2 : S ~b
}
$$
$$
\infer{ \vdash \arrow{arr}  ~\mathit{iso} : b ~R ~b}{
	 \vdash \mathit{iso} : b \leftrightarrow b
}
\quad
\infer{ \vdash r_1 \,\arrow{\ggg\xspace}\, r_2 : b_1 ~R ~b_3}{
	 \vdash r_1 : b_1 ~R ~b_2
	&
	 \vdash r_2 : b_2 ~R ~b_3
}
$$
$$
\infer{ \vdash \mathit{second} ~r : (b\times b_1) ~R ~(b\times b_2)}{
	 \vdash r : b_1 ~R ~b_2
}
\quad
\infer{ \vdash \mathit{state} ~s : 1 ~R ~b}{
	 \vdash s : S ~b
}
$$
$$
\infer{ \vdash \mathit{strength} : (b_1 \times S ~b_2) ~R ~(b_1\times b_2)}{
	 ~
}
$$
$$
\infer{ \vdash \eta _b : 1 ~R ~(b\times b)}{
	 ~
}
\quad
\infer{ \vdash \varepsilon  : (b\times b) ~R ~1}{
	 ~
}
$$
%subcode source qmll.tex:994

%%%%%%%%%%%%%%%%%%%%
\subsection{Semantics}

A program in \ensuremath{\mathcal{R}\Pi } consists of the application of a relation $r$ to a
set~$s$ producing a resulting set $s'$. We model this evaluation using two
applications: the application \ensuremath{r \,\overline{\,@\, }\, s} applies the relation \ensuremath{r} to the
set \ensuremath{s} and the application \ensuremath{r \,@\, v} applies the relation \ensuremath{r} to an
individual set element \ensuremath{v}. The first application~\ensuremath{\,\overline{\,@\, }\,} simply iterates
down the structure of the set until it finds individual elements which can be
processed using \ensuremath{\,@\,}:

\[\begin{array}{lcl }
 r \,\overline{\,@\, }\, \emptyset  &\mapsto & \emptyset  \\
 r \,\overline{\,@\, }\, \singleton {v} &\mapsto & r \,@\, v \\
 r \,\overline{\,@\, }\, (s_1 ~\uplus  ~s_2) &\mapsto & (r \,\overline{\,@\, }\, s_1) ~\uplus  ~(r \,\overline{\,@\, }\, s_2) \\
 \end{array}\]
%subcode source qmll.tex:1011

The application \ensuremath{\,@\,} applies a relation to a single value $v$ and returns
the set of possible values that are related to $v$:
\[\begin{array}{rclr}
 (\arrow{arr}  ~\mathit{iso}) \,@\, v &\mapsto & \singleton {\mathit{iso}(v)} \\
 (r_1 \,\arrow{\ggg\xspace}\, r_2) \,@\, v &\mapsto & r_2 \,\overline{\,@\, }\, (r_1 \,@\, v) \\
 (\mathit{second} ~r) \,@\, (v_1,v_2) &\mapsto & \mathit{strength} \,@\, (v_1, r \,@\, v_2) \\
 \mathit{strength} \,@\, (v,\emptyset ) &\mapsto & \emptyset  \\
 \mathit{strength} \,@\, (v_1,\singleton {v_2}) &\mapsto & \singleton {(v_1,v_2)} \\
 \mathit{strength} \,@\, (v,s_1 ~\uplus  ~s_2) &\mapsto & \mathit{strength} \,@\, (v,s_1) ~\uplus  \\
 && \mathit{strength} \,@\, (v,s_2) \\
 (\mathit{state} ~s) \,@\, () &\mapsto & s \\
 \eta _b \,@\, () &\mapsto & \biguplus_i  ~\singleton {(\mathit{v_i },\mathit{v_i })} ,\,\mathrm{\textbf{where} }\, \mathit{v_i } ~\!\!\in\!\!  ~b \\
 \varepsilon  \,@\, (v,v) &\mapsto & \singleton {()} \\
 \varepsilon  \,@\, (v,v') &\mapsto & \emptyset  ~\mathit{if} ~v \neq v' \\
 \end{array}\]
%subcode source qmll.tex:1026

When applied to a value \ensuremath{v}, the relation \ensuremath{\arrow{arr}  ~\mathit{iso}} simply applies the
underlying isomorphism to the value which is wrapped in a singleton
set. Applying the sequence \ensuremath{r_1} and \ensuremath{r_2} to a value \ensuremath{v} applies \ensuremath{r_1}
to \ensuremath{v} to produce a set which is passed to \ensuremath{r_2}. To apply a relation
\ensuremath{r} to the second component of a pair~\ensuremath{(v_1,v_2)}, we first produce
\ensuremath{(v_1,r\,@\, v_2)} and then use the tensorial strength of the monad to push the
pair construction through the resulting set as shown in the three rules for
\ensuremath{\mathit{strength}}. The relation \ensuremath{\mathit{state} ~s} maps the unit value \ensuremath{()} to the
given set \ensuremath{s}. When applied to the unit value \ensuremath{()}, the
relation~$\ensuremath{\eta }_b$ produces a superposition of all pairs of values $v_i$
where~$v_i$ is an element of the type $b$ (which is finite). The
relation~\ensuremath{\varepsilon } maps a pair of equal values to the singleton set
containing~\ensuremath{()} and maps a pair of different values to the empty set. As
the semantics for $\ensuremath{\eta }_b$ and \ensuremath{\varepsilon } shows, these constructs exploit
the fact that the underlying language of classical values is based on finite
values (and hence values that can be compared for equality and
enumerated). In order to accommodate sets whose elements are themselves sets
or relations, it is necessary to have a first-order representation of
relations using the arrow constructors instead of using the higher-order
Kleisli maps of the monad.

It is a routine task to verify that the evaluation rules preserve the types.
\begin{proposition}
If \ensuremath{\vdash r : b_1 ~R ~b_2} and \ensuremath{\vdash s : S ~b_1} then \ensuremath{\vdash r\,\overline{\,@\, }\, s : S ~b_2}.
\end{proposition}

%%%%%%%%%%%%%%%%%%%%
\subsection{Derived Relations} 
\label{derived-rel}

The language \ensuremath{\mathcal{R}\Pi } is surprisingly expressive: using standard categorical
constructions, it can express higher-order functions, currying, uncurrying,
recursion, adjoints, dot products, and outer products. We illustrate these
derived relations. The superdense coding protocol in
Section~\ref{sec:sdcoding} uses the \ensuremath{s_2r} construction below.

We start by defining for every relation \ensuremath{r} a relation \ensuremath{\mathit{first} ~r}
that applies \ensuremath{r} to the first component of a pair:
\[\begin{array}{rclr}
\mathit{first} &:& a ~R ~b \rightarrow (a \times c) ~R ~(b \times c) \\
\mathit{first} ~r &=& \arrow{arr}  \times_{\times } \,\arrow{\ggg\xspace}\, \mathit{second} ~r \,\arrow{\ggg\xspace}\, \arrow{arr}  \times_{\times } \\
 \end{array}\]
%subcode source qmll.tex:1068

Next we present currying and uncurrying and use them to turn any set of pairs
to a relation:
\[\begin{array}{rclr}
\mathit{curry} &:& ((c \times a) ~R ~b) \rightarrow (c ~R ~(a \times b)) \\
\mathit{curry} ~r &=& \arrow{arr}  ~(\curlyeqprec_{\times } \circ \times_{\times } ) \,\arrow{\ggg\xspace}\, \mathit{second} ~\eta _a \,\arrow{\ggg\xspace}\, \\
 && \arrow{arr}  \gtrless_{\times } \,\arrow{\ggg\xspace}\, \mathit{first} ~r \,\arrow{\ggg\xspace}\, \arrow{arr}  \times_{\times } \\
 \\
\mathit{uncurry} &:& (c ~R ~(a \times b)) \rightarrow ((c \times a) ~R ~b) \\
\mathit{uncurry} ~r &=& \mathit{first} ~r \,\arrow{\ggg\xspace}\, \arrow{arr}  ~(\times_{\times } \circ \gtrless_{\times } ) \,\arrow{\ggg\xspace}\, \\
 && \varepsilon  \,\arrow{\ggg\xspace}\, \curlyeqsucc_{\times } \\
 \\
s_2r &:& S ~(a,b) \rightarrow (a ~R ~b) \\
s_2r ~s  & = &  \arrow{arr}  \curlyeqprec_{\times } \,\arrow{\ggg\xspace}\, \mathit{uncurry} ~(\mathit{state} ~s) \\
 \end{array}\]
%subcode source qmll.tex:1082

We can also define a \emph{trace} operation to model recursion from cyclic
sharing~\cite{Hasegawa:1997:RCS:645893.671607}:
\[\begin{array}{rclr}
\mathit{trace} &:& ((a \times c) ~R ~(b \times c)) \rightarrow (a ~R ~b) \\
\mathit{trace} ~r &=& \arrow{arr}  \curlyeqprec_{\times } \,\arrow{\ggg\xspace}\, \mathit{first} ~\eta _c \,\arrow{\ggg\xspace}\, \\
 && \arrow{arr}  ~(\times_{\times } \circ \gtrless_{\times } ) \,\arrow{\ggg\xspace}\, \mathit{first} ~r \,\arrow{\ggg\xspace}\, \arrow{arr}  \lessgtr_{\times } \\
 && \,\arrow{\ggg\xspace}\, \mathit{second} ~\varepsilon  \,\arrow{\ggg\xspace}\, \arrow{arr}  ~(\times_{\times } \circ \curlyeqsucc_{\times } ) \\
 \end{array}\]
%subcode source qmll.tex:1090

Finally, we can also define an adjoint for every relation and use it to
define a \emph{costate} relation that matches a given state. We can combine a
state and a costate in two different ways to simulate the dot product and the
outer product constructions~\cite{categoricalQM}. 
\[\begin{array}{rclr}
\mathit{adjoint} &:& (a ~R ~b) \rightarrow (b ~R ~a) \\
\mathit{adjoint} ~r &=& \arrow{arr}  \curlyeqprec_{\times } \,\arrow{\ggg\xspace}\, \mathit{first} ~\eta _a \,\arrow{\ggg\xspace}\, \\
 && \mathit{first} ~(\mathit{second} ~r) \,\arrow{\ggg\xspace}\, \arrow{arr}  \lessgtr_{\times } \,\arrow{\ggg\xspace}\, \\
 && \mathit{second} ~\varepsilon  \,\arrow{\ggg\xspace}\, \arrow{arr}  ~(\times_{\times } \circ \curlyeqsucc_{\times } ) \\
 \\
\mathit{costate} &:& S ~a \rightarrow (a ~R ~1) \\
\mathit{costate} ~s  & = &  \mathit{adjoint} ~(\mathit{state} ~s) \\
 \\
\ip {.}{.} &:& S ~a \rightarrow S ~a \rightarrow (1 ~R ~1) \\
\ip {s_1}{s_2} &=& \mathit{state} ~s_1 \,\arrow{\ggg\xspace}\, \mathit{costate} ~s_2 \\
 \\
\op {.}{.} &:& S ~a \rightarrow S ~a \rightarrow (a ~R ~a) \\
\op {s_1}{s_2} &=& \mathit{costate} ~s_2 \,\arrow{\ggg\xspace}\, \mathit{state} ~s_1 \\
 \end{array}\]
%subcode source qmll.tex:1109
The result of the dot product is a relation of type {{1 R 1}} which
corresponds to a scalar~\cite{categoricalQM}.

%%%%%%%%%%%%%%%%%%%%
\subsection{Interpretations} 
\label{sec:interp}

There are two isomorphic ways of thinking about \ensuremath{\mathcal{R}\Pi }: a value of type 
\ensuremath{S ~b} could be viewed as a set of values of type~\ensuremath{b} or as a vector which
maps values of type \ensuremath{b} to scalars in the field of booleans. For example,
\ensuremath{s : S ~(\mathit{bool}  \times \mathit{bool} ) = \singleton {(\scalarZ ,\scalarZ )} \uplus  ~\singleton {(\scalarO ,\scalarO )}}
denotes the set \ensuremath{ \{ (\scalarZ ,\scalarZ ),(\scalarO ,\scalarO ) \}} in the first interpretation and
denotes the vector:
\[\begin{array}{r@{\!\!}l}
\begin{array}{r} 
(\scalarZ,\scalarZ) \\ (\scalarZ,\scalarO) \\ 
(\scalarO,\scalarZ) \\ (\scalarO,\scalarO) 
\end{array} & 
\begin{pmatrix}
\scalarO \\ \scalarZ \\ \scalarZ \\ \scalarO
\end{pmatrix}
\end{array}\]
in the second interpretation. In the vector notation, the labels on the left
enumerate all the values of the given type. A value is present if the
corresponding entry in the vector is $\scalarO$ and absent if the
corresponding entry is $\scalarZ$. To avoid clutter we will assume a fixed
preferred ordering of the labels on the left and omit them. 

Similarly a value of type \ensuremath{b_1 ~R ~b_2} can be viewed as a relation mapping
sets of values of type \ensuremath{b_1} to sets of values of type \ensuremath{b_2} or as a linear
map which given our preferred ordering could be represented as a matrix. For
example, the values:
\[\begin{array}{rclr}
r_1,r_2 &:& \mathit{bool}  ~R ~\mathit{bool}  \\
r_1 &=& s_2r ~(\singleton {(\scalarZ ,\scalarZ )} \uplus  ~\singleton {(\scalarO ,\scalarZ )} \uplus  ~\singleton {(\scalarO ,\scalarO )}) \\
r_2 &=& s_2r ~(\singleton {(\scalarZ ,\scalarZ )} \uplus  ~\singleton {(\scalarZ ,\scalarO )} \uplus  ~\singleton {(\scalarO ,\scalarZ )}) \\
 \end{array}\]
%subcode source qmll.tex:1145
denote the relations
$\{(\scalarZ,\scalarZ),(\scalarO,\scalarZ),(\scalarO,\scalarO)\}$ and
$\{(\scalarZ,\scalarZ),(\scalarZ,\scalarO),(\scalarO,\scalarZ)\}$ in the
first interpretation, and the matrices:
\[
\ensuremath{r_1 =} \begin{pmatrix}
\scalarO & \scalarO \\
\scalarZ & \scalarO
\end{pmatrix}
\qquad
\ensuremath{r_2 =} \begin{pmatrix}
\scalarO & \scalarO \\
\scalarO & \scalarZ
\end{pmatrix}
\]
in the second interpretation. In the matrix notation, the columns are
implicitly indexed with $\scalarZ$ and $\scalarO$ from left to right, and the
rows are implicitly indexed with $\scalarZ$ and $\scalarO$ from top to
bottom. An entry is $\scalarO$ if the pair
$(\textrm{column-label},\textrm{row-label})$ is in the relation and
$\scalarZ$ otherwise.

In a conventional setting, the composition of $r_1$ and $r_2$ would be:
\[ \ensuremath{r_1 \,\arrow{\ggg\xspace}\, r_2} =  \{ (\scalarZ,\scalarZ),(\scalarZ,\scalarO),
  (\scalarO,\scalarZ),(\scalarO,\scalarO) \} 
\] 
This is the semantics one gets in conventional relational programming, i.e.,
pure Prolog and backtracking
monads~\cite{Hinze:deriving,Kiselyov:2005:BIT:1086365.1086390} which indeed
can be implemented and formalized using sequences or sets~\cite{Wadler85}.

However in \ensuremath{\mathcal{R}\Pi }, the composition of these two same relations is $\{
(\scalarZ,\scalarZ),(\scalarZ,\scalarO),(\scalarO,\scalarO) \}$ because the
pair $(\scalarO,\scalarZ)$ can be produced in two different ways which cancel
due to interference. It is perhaps more intuitive to compute the composition
in the world of matrices:
\[
\ensuremath{r_1 \,\arrow{\ggg\xspace}\, r_2} = 
\begin{pmatrix} 
\scalarO & \scalarO \\
\scalarO & \scalarZ
\end{pmatrix} 
\begin{pmatrix} 
\scalarO & \scalarO \\
\scalarZ & \scalarO
\end{pmatrix} =
\begin{pmatrix} 
\scalarO \scalarPlus \scalarZ & \scalarO \scalarPlus \scalarO \\
\scalarO \scalarPlus \scalarZ & \scalarO \scalarPlus \scalarZ
\end{pmatrix} =
\begin{pmatrix} 
\scalarO & \scalarZ \\
\scalarO & \scalarO 
\end{pmatrix} 
\]

We can give similar interpretations for each of the \ensuremath{\mathcal{R}\Pi } constructs. In
the world of matrices, \ensuremath{\mathit{second} ~r} denotes the matrix produced by the tensor
product of the identity matrix and the matrix corresponding to \ensuremath{r}. The construct
\ensuremath{\mathit{state}} is a no-op: it allows us to view a vector of size \ensuremath{n} as a matrix
of dimensions \ensuremath{n \times 1}. We illustrate the matrix denoted by \ensuremath{\mathit{strength}} at
the type \ensuremath{ (\mathit{bool}  \times S ~\mathit{bool} ) ~R ~(\mathit{bool}  \times \mathit{bool} )}
\[
\begin{pmatrix}
\ensuremath{\scalarZ } & \ensuremath{\scalarO } & \ensuremath{\scalarZ } & \ensuremath{\scalarO } & \ensuremath{\scalarZ } & \ensuremath{\scalarZ } & \ensuremath{\scalarZ } & \ensuremath{\scalarZ } \\
\ensuremath{\scalarZ } & \ensuremath{\scalarZ } & \ensuremath{\scalarO } & \ensuremath{\scalarO } & \ensuremath{\scalarZ } & \ensuremath{\scalarZ } & \ensuremath{\scalarZ } & \ensuremath{\scalarZ } \\
\ensuremath{\scalarZ } & \ensuremath{\scalarZ } & \ensuremath{\scalarZ } & \ensuremath{\scalarZ } & \ensuremath{\scalarZ } & \ensuremath{\scalarO } & \ensuremath{\scalarZ } & \ensuremath{\scalarO } \\
\ensuremath{\scalarZ } & \ensuremath{\scalarZ } & \ensuremath{\scalarZ } & \ensuremath{\scalarZ } & \ensuremath{\scalarZ } & \ensuremath{\scalarZ } & \ensuremath{\scalarO } & \ensuremath{\scalarO } 
\end{pmatrix}
\]
The rows are indexed from top to bottom by \ensuremath{(\scalarZ ,\scalarZ )}, \ensuremath{(\scalarZ ,\scalarO )}, \ensuremath{(\scalarO ,\scalarZ )},
and \ensuremath{(\scalarO ,\scalarO )}. The columns are indexed from left to right by 
\ensuremath{(\scalarZ ,\emptyset )},
\ensuremath{(\scalarZ ,\singleton {\scalarZ })}, 
\ensuremath{(\scalarZ ,\singleton {\scalarO })}, 
\ensuremath{(\scalarZ ,\singleton {\scalarZ } \uplus  ~\singleton {\scalarO })}, 
\ensuremath{(\scalarO ,\emptyset )},
\ensuremath{(\scalarO ,\singleton {\scalarZ })}, 
\ensuremath{(\scalarO ,\singleton {\scalarO })}, and
\ensuremath{(\scalarO ,\singleton {\scalarZ } \uplus  ~\singleton {\scalarO })}.

The matrices corresponding to \ensuremath{\eta _b} and \ensuremath{\varepsilon } are column and row
vectors with entries \ensuremath{\scalarO } at the diagonal elements. For example, 
\ensuremath{\eta _\mathit{bool}  : 1 ~R ~(\mathit{bool}  \times \mathit{bool} )} and  \ensuremath{ \varepsilon  : (\mathit{bool}  \times \mathit{bool} ) ~R ~1} are:
\[
\ensuremath{\eta } = \begin{pmatrix}
\scalarO \\
\scalarZ \\
\scalarZ \\ 
\scalarO
\end{pmatrix}
\qquad
\ensuremath{\varepsilon } = \begin{pmatrix}
\scalarO & 
\scalarZ &
\scalarZ & 
\scalarO
\end{pmatrix}
\]

As a special case, we note that relations of type \ensuremath{1 ~R ~1} denote matrices
of dimension \ensuremath{1 \times 1}, i.e., scalars. We interpret the matrix
$\begin{pmatrix}\scalarO\end{pmatrix}$ as the scalar $\scalarO$ and the
matrix $\begin{pmatrix}\scalarZ\end{pmatrix}$ as the scalar~$\scalarZ$.

\begin{lemma}
\label{seq-unfold}
The semantics of \ensuremath{\mathcal{R}\Pi } is \emph{sound} with respect to both the relational
interpretation and the vector interpretation.
\end{lemma}

%%%%%%%%%%%%%%%%%%%%
\subsection{Dagger Compact Closed Structure} 

We can now verify that the arrow fragment of the language \ensuremath{\mathcal{R}\Pi } forms a
dagger compact closed category. The objects of the category are the types 
\ensuremath{S ~b} and the morphisms are the relations \ensuremath{b_1 ~R ~b_2}. These form a category
since we have identity morphisms due to \ensuremath{\arrow{arr} ~\mathit{id}} and the composition of
relations is associative due to the associativity of the matrix
multiplication in the model (soundness).  The category has the required
monoidal structure with \ensuremath{S ~(b_1 \times b_2)} being the product for the objects 
\ensuremath{S ~b_1} and \ensuremath{S ~b_2} and \ensuremath{S ~1} is the identity object for tensors and the
required natural transforms $\alpha$, $\lambda$ and $\rho$ are \ensuremath{\arrow{arr} } lifted
forms of their \ensuremath{\Pi } equivalents. Similarly braiding and symmetry are
preserved by lifting \ensuremath{\times_{\times }} from \ensuremath{\Pi }.

The dagger structure on relations is given by the contravariant
functor \ensuremath{\mathit{adjoint} : b_1 ~R ~b_2 \rightarrow b_2 ~R ~b_1} which we constructed in
Section \ref{derived-rel} and is easily verified to be involutive.

Finally the required autonomous structure is given by choosing the dual of
\ensuremath{S ~b} to be the object \ensuremath{S ~b} itself. The required {\it unit} and 
{\it counit} are given by \ensuremath{\eta } and \ensuremath{\varepsilon } in \ensuremath{\mathcal{R}\Pi }. The required
adjunction triangles become the same as the right and left duals on objects
coincide. Thus the arrow fragment of \ensuremath{\mathcal{R}\Pi } forms a dagger compact closed
category.

\section{Using \ensuremath{\mathcal{R}\Pi } for Quantum Computation}

Two more things are needed to use \ensuremath{\mathcal{R}\Pi } for quantum computation. First, the
zero vector (which is denoted by \ensuremath{\emptyset }) should never be encountered
during computation as it does not correspond to a valid quantum
state. Second, we must define a notion of measurement.

%%%%%%%%%%%%%%%
\subsection{Evolution of Quantum States} 

As Schumacher and Westmoreland~\cite{modalqm} explain, the evolution of
quantum states in \DQC\ only requires that the linear maps be invertible. We
have already established that every linear map has an adjoint but these
adjoints do not necessarily coincide with the inverses. Consider the matrix:
\[
\begin{pmatrix}
\ensuremath{\scalarO } & \ensuremath{\scalarO } \\
\ensuremath{\scalarO } & \ensuremath{\scalarO } 
\end{pmatrix}
\]
The adjoint of this matrix is itself but when multiplied by itself in the
field of booleans it produces the zero matrix. In other words, any \ensuremath{\mathcal{R}\Pi }
expression that can denote this matrix should not be allowed. All the
expressions in the syntactic category \ensuremath{r} in \ensuremath{\mathcal{R}\Pi } denote invertible
matrices except for \ensuremath{\varepsilon }. This suggests a possibility for tracking the
uses of \ensuremath{\varepsilon } using an extended type system to syntactically guarantee
that a given expression denotes an invertible transformation.

\subsection{Measurement} 

In standard QM, the postulate of measurement requires the notion of an
\emph{orthogonal projection}. However, as explained in Section~\ref{sec:dqt},
discrete quantum theories lack inner products and hence lack a standard
notion of orthogonality. Nevertheless it is possible to define a sensible
notion of measurement as explained by Schumacher and
Westmoreland~\cite{modalqm}. The key technical observations are (i) any set
of linearly independent vectors can form a basis, and (ii) it is possible to
define dual vectors for a given basis.

In more detail, assuming the vector interpretation of \ensuremath{\mathcal{R}\Pi }, a measurement
(or an observable) corresponds to a basis $\{s_1,s_2,\ldots\}$ where the
collection of values $s_i$ denote linearly independent vectors. As discussed
in Section~\ref{sec:dqt}, the space of 1-qubit vectors contains just three
vectors with any pair forming a basis. For each choice of basis, we associate
an observable:
\[\begin{array}{rcl}
X\textit{-basis} &=& 
  \{ \ensuremath{\singleton {\scalarO }, \singleton {\scalarZ } \uplus  ~\singleton {\scalarO }} \} \\
Y\textit{-basis} &=& 
  \{ \ensuremath{ \singleton {\scalarZ } \uplus  ~\singleton {\scalarO }, \singleton {\scalarZ }} \} \\
Z\textit{-basis} &=& 
  \{ \ensuremath{\singleton {\scalarZ }, \singleton {\scalarO }} \}
\end{array}\]
We then associate a \emph{basis dependent} dual $\overline{s}_i$ to each
vector $s_i$ such that \ensuremath{ \ip {\overline {\mathit{s_i }} }{\mathit{s_j }}} denotes the scalar
$\scalarO$ if and only if $i=j$. For example, the dual vectors for the
vectors in the $X\textit{-basis}$ are:
\[\begin{array}{rcl}
X\textit{-dual}_X &=& 
  \{ \ensuremath{ \singleton {\scalarZ } \uplus  ~\singleton {\scalarO }, \singleton {\scalarZ }} \} 
\end{array}\]

The simplest way then to extend \ensuremath{\mathcal{R}\Pi } is to add one construct \ensuremath{\mathit{measure}}
that takes two arguments: a vector and a set of dual basis vectors along
which the vector should be measured. For example, to measure the vector 
\ensuremath{\singleton {\scalarO }} in the $X\textit{-basis}$, we would write: 
\ensuremath{ \mathit{measure} ~\singleton {\scalarO } \{ \singleton {\scalarZ } \uplus  ~\singleton {\scalarO }, \singleton {\scalarZ } \}}
The result of measurement is any dual basis vector that can possibly match the
given vector selected at random. In the above example, only the first dual
basis vector can possibly match the given vector which means that the result
is deterministic. 

A remarkable feature of the categorical approach to QM is that the semantics
of measurement can be expressed as follows in \ensuremath{\mathcal{R}\Pi }:

\[
\ensuremath{\mathit{measure} ~s ~\mathit{~ } \{ d_1, d_2, ... \}} = d_i ~\mbox{if}~
  \ensuremath{ \ip {\mathit{d_i }}{s}} ~\mbox{denotes}~ \begin{pmatrix}\scalarO\end{pmatrix}
\]
As explained in Sections~\ref{derived-rel} and~\ref{sec:interp}, the dot
product is expressible in \ensuremath{\mathcal{R}\Pi } and the relations of type \ensuremath{1 ~R ~1} are
identified with scalars. If more than one dual vector produces the scalar $\scalarO$
then the result of measurement is non-deterministic with the 
dual vector $d_i$ picked at random. 

It should also be possible to exploit the fact that our categories have
biproducts to extend the language in a richer way by allowing measurements to
occur in the middle of computation~\cite{categoricalQM}. 

\subsection{Example: Superdense Coding} 
\label{sec:sdcoding}

The superdense coding example presented by Schumacher and
Westmoreland~\cite{modalqm} can be directly implemented in \ensuremath{\mathcal{R}\Pi }. Alice and
Bob initially share an entangled state, represented by 
\ensuremath{\singleton {(\scalarZ , \scalarZ )} \uplus  ~\singleton {(\scalarO , \scalarO )}}.  
Depending on Alice's choice of sending numbers 0
to 3, Alice applies one of the following operations on the first bit:
\[\begin{array}{lcl }
 \mathit{alice}_0 = \arrow{arr}  ~\mathit{id} \\
 \mathit{alice}_1 = \arrow{arr}  ~(\inplus  \circ \times_{\times } \circ \niplus ) \\
 \mathit{alice}_2 = s_2r ~(\mathit{state} ~(\singleton {(\scalarZ , \scalarZ )} \uplus  ~\singleton {(\scalarO , \scalarZ )} \uplus  ~\singleton {(\scalarO , \scalarO )})) \\
 \mathit{alice}_3 = \mathit{alice}_1 \,\arrow{\ggg\xspace}\, \mathit{alice}_2 \\
 \end{array}\]
%subcode source qmll.tex:1660

By measuring in a particular dual basis described below, Bob will
deterministically set a different dual vector for each possible operation that
Alice could have performed. The entire example is then written as follows:
\[\begin{array}{rclr}
 \mathit{measure} ~((\mathit{first} ~\mathit{alice}_n) \,\overline{\,@\, }\, (\singleton {(\scalarZ , \scalarZ )} \uplus  ~\singleton {(\scalarO , \scalarO )})) ~\mathit{dualbasis} \\
 \end{array}\]
%subcode source qmll.tex:1666

\[\begin{array}{ll }
 \textbf{where}  ~\mathit{dualbasis} = \{ & \singleton {(\scalarZ ,\scalarO )} \uplus  ~\singleton {(\scalarO ,\scalarZ )} \uplus  ~\singleton {(\scalarO ,\scalarO )}, \\
 & \singleton {(\scalarZ ,\scalarZ )} \uplus  ~\singleton {(\scalarO ,\scalarZ )} \uplus  ~\singleton {(\scalarO ,\scalarO )}, \\
 & \singleton {(\scalarZ ,\scalarO )} \uplus  ~\singleton {(\scalarO ,\scalarZ )}, \\
 & \singleton {(\scalarZ ,\scalarZ )} \uplus  ~\singleton {(\scalarO ,\scalarO )} \} \\
 \end{array}\]
%subcode source qmll.tex:1673

Interestingly, it is possible to write this same example in conventional
relational programming. The fact that intermediate results are accumulated
using the standard union instead of the exclusive union is apparent in the
results. However because this computation does not use significant
intermediate steps, the only values that need to be treated specially are the
final results. Indeed if the results that appear an even number of times are
removed, then the Prolog execution performs the superdense coding exactly. 

The complete program and its execution are below:
\begin{multicols}{2}
\setlength{\columnseprule}{1pt}
\lstset{
        language=Prolog,
 morekeywords={true, false},
	keywordstyle=\bfseries\ttfamily,
	identifierstyle=\ttfamily,
	commentstyle=\ttfamily,
	stringstyle=\ttfamily,
	showstringspaces=false,
	basicstyle=\scriptsize\sffamily,
	tabsize=2,
	breaklines=true,
	prebreak = \raisebox{0ex}[0ex][0ex]{\ensuremath{\hookleftarrow}},
	breakatwhitespace=false,
	aboveskip={1.2\baselineskip},
        columns=fixed,
        extendedchars=true,
}
\begin{lstlisting}
r(false,false).
r(true,true).

s(false,true).
s(true,false).

u(false,false).
u(false,true).
u(true,true).

v(false,false).
v(false,true).
v(true,false).

rd(false,true).
rd(true,false).
rd(true,true).

sd(false,false).
sd(true,false).
sd(true,true).

ud(false,true).
ud(true,false).

vd(false,false).
vd(true,true).

eq(false,false).
eq(true,true).

id(false,false).
id(true,true).

g(false,true).
g(true,false).

k(false,false).
k(true,false).
k(true,true).
 \end{lstlisting}
%subcode source qmll.tex:1727

\lstset{
        language=Prolog,
 morekeywords={true, false},
	keywordstyle=\bfseries\ttfamily,
	identifierstyle=\ttfamily,
	commentstyle=\ttfamily,
	stringstyle=\ttfamily,
	showstringspaces=false,
	basicstyle=\scriptsize\sffamily,
	tabsize=2,
	breaklines=true,
	prebreak = \raisebox{0ex}[0ex][0ex]{\ensuremath{\hookleftarrow}},
	breakatwhitespace=false,
	aboveskip={1.2\baselineskip},
        columns=fixed,
        extendedchars=true,
}
\begin{lstlisting}
gk(X,Y) :- g(X,Z),k(Z,Y).

alice(0,X,Y):- id(X,Y).
alice(1,X,Y):- g(X,Y).
alice(2,X,Y):- k(X,Y).
alice(3,X,Y):- gk(X,Y).
 \end{lstlisting}
%subcode source qmll.tex:1736
\end{multicols}

\vspace{-20pt}
\lstset{
        language=Prolog,
 morekeywords={true, false},
	keywordstyle=\bfseries\ttfamily,
	identifierstyle=\ttfamily,
	commentstyle=\ttfamily,
	stringstyle=\ttfamily,
	showstringspaces=false,
	basicstyle=\scriptsize\sffamily,
	tabsize=2,
	breaklines=true,
	prebreak = \raisebox{0ex}[0ex][0ex]{\ensuremath{\hookleftarrow}},
	breakatwhitespace=false,
	aboveskip={1.2\baselineskip},
        columns=fixed,
        extendedchars=true,
}
\begin{lstlisting}
sdcoding(N,M) :- r(X,Y),alice(N,X,B),measure((B,Y),M).

measure((S1,S2),0) :- rd(B1,B2),dotP((B1,B2),(S1,S2)).
measure((S1,S2),1) :- sd(B1,B2),dotP((B1,B2),(S1,S2)).
measure((S1,S2),2) :- ud(B1,B2),dotP((B1,B2),(S1,S2)).
measure((S1,S2),3) :- vd(B1,B2),dotP((B1,B2),(S1,S2)).

dotP((B1,B2),(S1,S2)) :- eq(B1,S1),eq(B2,S2).
 \end{lstlisting}
%subcode source qmll.tex:1749

\begin{multicols}{2}
\setlength{\columnseprule}{1pt}
{\scriptsize\begin{verbatim}
28 ?- sdcoding(0,X).
X = 1 ;
X = 3 ;
X = 0 ;
X = 1 ;
X = 3.

29 ?- sdcoding(1,X).
X = 0 ;
X = 1 ;
X = 2 ;
X = 0 ;
X = 2.
\end{verbatim}
% \columnbreak
\begin{verbatim}
30 ?- sdcoding(2,X).
X = 1 ;
X = 3 ;
X = 0 ;
X = 2 ;
X = 0 ;
X = 1 ;
X = 3.

31 ?- sdcoding(3,X).
X = 1 ;
X = 3 ;
X = 0 ;
X = 1 ;
X = 2 ;
X = 0 ;
X = 2.
\end{verbatim}}
\end{multicols}

%%%%%%%%%%%%%%%%%%%%%%%%%%%%%%%%%%%%%%%%%%%%%%%%%%%%%%%%%%%%%%%%%%%%%%%%%%%%%
\section{Conclusion}
\label{sec:conc} 

We have distilled a programming model for the discrete quantum theory
over the field of booleans recently introduced by Schumacher and
Westmoreland~\cite{modalqm}. The model is expressed in a small
calculus \ensuremath{\mathcal{R}\Pi } with formal type rules and semantics. The language
\ensuremath{\mathcal{R}\Pi } is directly inspired by the computational structures of
quantum mechanics previously identified by Abramsky and Coecke and
Selinger. The semantics of \ensuremath{\mathcal{R}\Pi } is sound with respect to both a
relational model with exclusive unions or a vector space model over
the field of booleans. Computationally \ensuremath{\mathcal{R}\Pi } is a relational
programming language surprisingly similar to traditional relational
programming langauges like Prolog with the significant difference that
where Prolog would accumulate all possible answers with repetitions,
\ensuremath{\mathcal{R}\Pi } allows interference between possible answers.

A natural goal of our research is to unravel the mathematical
underpinnings behind the power of quantum computation. By analyzing
models such as those presented in this paper, we hope we will be able
to shed some light on this fundamental issue. Deutsch's quantum
algorithm, which establishes in a single measurement whether a boolean
function is constant or balanced, makes use of interference and
quantum parallelism. Although discrete quantum computing over the
field of booleans, \DQC, contains both these properties, Deutsch's
algorithm cannot be efficiently implemented within this model. This
example illustrates the non-trivial character of the quest to
disentangle the power of quantum computation. In more detail, having
eliminated much of the structure of actual quantum mechanics, the
connection between \DQC~ and conventional relational programming
singles out the use of exclusive union as the source of all the
``quantum-ness''. Assuming that the exclusive union would be performed
in ``constant time'' by a quantum computer, more work is needed to
establish whether it is possible to write more efficient algorithms
using the exclusive union.

\section*{Acknowledgments}
We have benefited from many discussions with the Quantum and Natural
Computing group at Indiana University, especially Mike Dunn, Andy Hanson, and
Larry Moss. The presentation was improved following suggestions from Daniel
P. Friedman.

%%\IEEEtriggeratref{8}

\bibliographystyle{IEEEtran}
\bibliography{IEEEabrv,p}
\end{document}